\documentstyle[prl,aps]{revtex}

\begin{document}

\draft

\title{New constraints from Haverah Park data on the photon and iron fluxes of UHE cosmic rays}

\author{M. Ave$^1$, J.A.~Hinton$^2$, R.A.~V\'azquez$^1$, 
A.A.~Watson$^2$, and E.~Zas$^1$\\}
\address{$^{1}$ Departamento de F\'\i sica de Part\'\i culas,
Universidad de Santiago, 15706 Santiago de Compostela, Spain. \\
$^{2}$ Department of Physics and Astronomy, University of Leeds, Leeds LS2 9JT ,UK \\} 

\date{\today}
\maketitle
\begin{abstract}
Using data from inclined events ($60^{\circ}<\theta<80^{\circ}$)
recorded by the Haverah Park shower detector, we show that above
$10^{19}$ eV less than $30\%$ of the primary cosmic rays can be
photons or iron nuclei at the $95 \%$ confidence level. Above 
$4 \times 10^{19}$ eV less than $55\%$
of the cosmic rays can be photonic at the same confidence level. 
These limits place important
constraints on some models of the origin of ultra high energy cosmic rays.
Details of two new events above $10^{20}$~eV are reported.
\end{abstract}

The highest energy cosmic rays above the Greisen-Zatsepin-Kuzmin (GZK)
cut-off \cite{GZK} are a mystery both in terms of their origin and
their mass composition. Conventional acceleration mechanisms,
so called `bottom up' scenarios, predict an
extragalactic origin with mainly proton composition as, although nuclei
of higher charge are more easily accelerated, they are fragile to
photonuclear processes in the strong photon fields to be expected in
likely source regions \cite{conventional}. ``Top down'' models 
explain the highest energy cosmic
rays as arising from the decay of some sufficiently massive ``X-particles''.
These models predict particles such as nucleons, photons and even possibly 
neutrinos as the high energy cosmic rays, but not heavy nuclei. 
In some models \cite{Berezinsky,Sarkar} these X-particles are postulated
as long-living metastable super-heavy relic particles (MSRP) clustering in
our galactic halo.  For these MSRP models a photon dominated primary 
composition at $10^{19}$ eV is expected. Other top down models \cite{TD} associate 
X-particles with processes involving systems of cosmic topological defects
which are uniformly distributed in the universe, and predict a photon dominated
composition only above $\sim 10^{20}$~eV. These models are affected by the constraint that 
the low energy photons ($\sim$100~MeV) arising from interactions of UHE photons with the cosmic
microwave background cannot be larger than the observed diffuse low
energy flux \cite{Chi}. Observations above $10^{19}$~eV are currently consistent with both interpretations
\cite{FlysEye,AGASA}. There is however some partial evidence against
the photon hypothesis. Shower development of the highest energy event
\cite{FlysEye}, is inconsistent with a photon initiated shower
\cite{halzen} while AGASA measurements of the muon lateral
distribution of the highest energy events are compatible with a proton
origin~\cite{agasacorsika}. No measurement of, or limit to, the photon flux above
$10^{19}$~eV has been reported.

Here we describe a new method which we use to set a limit to the photon and iron content
of the highest energy cosmic rays. We show that observations of
inclined showers provide a powerful tool to discriminate between
photon and hadron dominated compositions. For primaries arriving at
zenith angles, $\theta$ $>60^{\circ}$ the shower particles reaching
sea-level are almost entirely muons, with a small contamination of
electrons and gammas arising mainly from muon decay~\cite{rate}.  From
our simulations we find that photons are expected to produce fewer
muons than hadrons (a factor $\sim$9) at $10^{19}$~eV. This factor
decreases with shower energy because of the rise of the
photoproduction cross section and the decrease of the pair production
and bremsstrahlung cross sections (due to LPM suppression~\cite{LPM}).
Our conclusions on the photon flux are not sensitive to the choice of
model: the implementation of photohadronic interactions in the AIRES
code~\cite{AIRES} and CORSIKA code~\cite{CORSIKA} (using the
parameterization of \cite{Stanev}) give predictions of the total muon
number that are equal to within 10\% at $10^{20}$~eV. In addition our
simulations show that the shape of the lateral distribution of muons
in inclined showers is constant with energy and is insensitive to
shower to shower fluctuations in longitudinal development~\cite{rate}.

Here we use data from the Haverah Park array, a 12~km$^{2}$ array of
water-\v Cerenkov detectors~\cite{Haverah}. The data used were
recorded between 1974 and 1987 and comprise around 8000 events with
$\theta > 60^{\circ}$ from an on-time of $3.6 \times 10^{8}$~s. These
events were not analysed originally because the limited computing
power then available required assumptions of circular symmetry which
are not valid for inclined showers due to geomagnetic field effects.
The analysis described below yielded 46 events with $E > 10^{19}$ eV.
Seven events have energies $> 4 \times 10^{19}$~eV and two events have
energies $>10^{20}$ eV. We show that the rate observed for inclined
showers is consistent with a proton dominated primary composition and
significantly above that expected if the primary composition is
dominated by photons.

Inclined showers recorded at Haverah Park have been analysed for
direction and energy using a combination of Monte Carlo techniques and
muon density parameterizations, see \cite{model,rate} for details.
For zenith angles in the range $60^\circ$~-~$89^\circ$ (in $1^\circ$
steps) muon density maps were generated using the model~\cite{model}
with inputs from AIRES for the QGSJET hadronic model \cite{QGSJET}
with $10^{19}$~eV protons.  Different azimuth angles are modeled as
described in \cite{model}. Throughout we assume the representation of
the energy spectrum recently given in \cite{NW}.  We find that the
lateral and energy distributions of muons in inclined showers are
largely independent of primary composition and energy so that
simulation of different primary energies and compositions is achieved
by scaling the muon density maps described above. We find $N_{\mu}
\propto E^{\alpha}$ with $\alpha$ equal to 0.924, 0.906 and 1.20 for
proton, iron and gamma primaries respectively. The relative total muon
numbers at $10^{19}$~eV are 1.0, 1.36, 0.11 for proton, iron and gamma
primaries respectively. In general differences between the
lateral distributions and energy spectra of muons in photon and
hadronic showers are of particular importance.  For inclined
showers, however, the differences 
decrease as the zenith angle increases 
because the mean height of muon production also increases
for both types of primary. At 60$^\circ$,
the lowest zenith angle considered,
the differences in the constant density contours
are below 20 $\%$ for distances between 300 and 1000 m. 
We adopt parameters 
appropriate to proton primaries to give a conservative estimate of 
the shower energies by comparison with what would be derived from 
the assumption of gamma ray primaries. By fitting density 
maps for proton primaries
on an event by event basis we thus obtain equivalent proton energies 
$E_p$. For other primaries
the energy is related to $E_p$ by an energy dependent multiplicative
factor which is $\sim$6 (0.7) for gamma (iron) primaries at $E_{p} =
10^{19}$~eV. i.e. a photon would require an energy 6 times that of a
proton to produce a given density map.

In addition to the electromagnetic contribution due to muon decay,
which is present at all core distances at the $20 \%$ level for these
detectors, the tail of the electromagnetic part of the shower is
important at zenith angles below $70^{\circ}$ and core distances less
than 500~m.  This contribution is modeled using AIRES with QGSJET and
is radially symmetric in the shower plane. The tail of the
electromagnetic part of the shower contributes $10 \%$ of the total
water-\v Cerenkov signal at 500~m from the core for a $60^{\circ}$
shower.

The Haverah Park arrival directions were determined originally using
only the 4 central triggering detectors \cite{rate}. We have
reanalysed the arrival directions of showers having original values of
$\theta$ $>56^\circ$, taking into account all detectors which have
timing information. This reanalysis produces smaller arrival direction
uncertainties.  The rate, as a function of zenith angle, obtained with
the new zenith angles, is consistent with predictions \cite{rate}
showing that the zenith angle reconstruction and the response of the
array to inclined showers are well understood.

The curvature of the shower front has been investigated using the
AIRES code for inclined showers and found to be consistent with the
simple approximation of a spherical front centered on the mean
production height of the muons (e.g.  at 60$^{\circ}$ the radius of
curvature is 16 km \cite{model}).  Beyond $\sim 80^{\circ}$ curvature
effects are rather small and it is usually sufficient to assume a
plane front~\cite{billior}. When the detected muon number is small
there is a systematic effect on the curvature correction and large
fluctuations due to limited sampling of the shower front. Therefore,
we disregard the timing information from detectors with $<15$ detected
equivalent muons. Because of the dependence of the curvature fit on
the position of the shower core a three step iteration was needed to give
convergence of the core location and direction fits.

The detector signals were measured in units of vertical equivalent
muons. Using the GEANT based package, WTANK \cite{WTANK}, we find that
this unit corresponds to an average number of 14 photoelectrons, 
consistent with an early experimental estimate of 15 photoelectrons
\cite{Baxter}. For
inclined showers additional effects, such as direct light on the
photomultiplier tubes, delta rays, and pair production and
bremsstrahlung by muons inside the tank, increase this number. For a given
zenith angle, the recorded signals are converted into the number of
photoelectrons and hence the muon density. 
The simulations take full account of stopping muons and the 
resulting decay electrons.

The observed densities were fitted against predictions using the maximum
likehood method.  Poissonian errors, measurement errors and errors
due to the uncertainty in detector geometry were included. Some
events contain saturated detectors which were accounted for using a
gaussian integral for the likelihood function.  A three dimensional
grid search was made to find the impact point and energy of the
shower. The energy was varied in the range $10^{17} < E_p <
10^{21}$~eV in steps of $0.1$ in $\log_{10}(E_p/\mbox{eV})$. The
impact point was varied over a grid of 12 km $\times$ 6 km in 40 m
steps in the perpendicular plane, the grid asymmetry being necessary
to accommodate the ellipticity of inclined showers.

The photoelectron distributions from a water detector show long tails
due to the processes mentioned above \cite{rate}.  We therefore expect
an excess of upward over downward fluctuations from the average
detector signal.  For each event the number of deviations
$>$2.5$~\sigma$ expected is calculated from the expected photoelectron
distributions. We reject signals having (upward or downward)
deviations greater than 2.5~$\sigma$, recalculating the best fit core
after any rejection.  Of 211 densities in the events of
table~\ref{tab} we rejected 13 upward deviations (the expected number
was 17) and rejected 4 densities with downward deviations $>2.5
\sigma$.

Errors in the energy and core determination were determined from the
likelihood function as in \cite{lampton}.  In addition to this
error, an error in energy arises due to the uncertainty in the zenith
angle. The error from the zenith angle determination and the error from the
fit for core and energy are added in quadrature to give the total
error shown in table \ref{tab}.  To guarantee the quality of events
the following cuts were made: (i) the distance from the central
triggering detector to the core position in the shower plane $<2$~km,
(ii) $\chi^2$ probability for the energy and direction fits $>$ 1 \%,
(iii) the downward error in the energy determination be less than a
factor of 2.  For $>80^{\circ}$ no showers pass cut (iii).

In figure \ref{events.fig} are density maps for two events. These are
plotted in the plane perpendicular to the shower direction together
with the contours of densities that best fit the data. In each figure
the array is rotated in the shower plane such that the y-axis is
aligned with the component of the magnetic field perpendicular to the
shower axis.  In figure 1b the asymmetry in the density pattern due to
the geomagnetic field is apparent.  For both events the core is
surrounded by recorded densities and is well determined.  In table
\ref{tab} details are given for 7 events with $E_{p} > 4 \times
10^{19}$ eV.

The data described above are compared to the result expected from
different primaries using an input energy spectrum \cite{NW} and a
Monte Carlo calculation. Figure \ref{rate.fig} shows the resulting
spectra, for three primary compositions, compared to the data. These
simulated spectra are somewhat dependent on the high energy
interaction model used.  The result is shown for the AIRES air-shower
code with the QGSJET interaction model. The SIBYLL hadronic
interaction model \cite{Sibyll} produces fewer muons than QGSJET (36\%
less at 10$^{19}$ eV) resulting in reconstructed energies that are
higher by $\approx$40\%. Using spectra from the QGSJET model we find
that less than $30 \%$ of primary cosmic rays above $10^{19}$ eV can
be iron, at a $95 \%$ confidence level (assuming a two component mass
composition).

The ratio of photons to protons for MSRP models was first given as 
typically 10\cite{Berezinsky} at 10$^{19}$ eV. However a later model 
predicts a ratio closer to 2 \cite{Sarkar}. On general grounds 
dominance of photons over protons is expected for these models due to the QCD
fragmentation functions of X-particles to mesons and baryons. From
figure 2 we deduce that above $10^{19}$ eV less than $30 \%$ of the
primary cosmic rays can be photons, with a $95 \%$ confidence level.
Above $4 \times 10^{19}$ eV less than $55 \%$ can be photons at the
same confidence level. Here we have assumed that downward
or upward poissonian fluctuations from the observed integral numbers
by 2 standard deviations could be accounted for by appropriate
contributions of protons plus gamma rays or protons plus iron nuclei
respectively. 

These limits set important constraints to TD mechanisms as the origin
of the highest energy cosmic rays.  We note also that the
gamma/proton ratio predicted to arise from proton interactions with
the 2.7 K background radiation is $30\%$ at $10^{19}$ eV when it is 
assumed
that the protons are produced universally with a differential
slope of 2 and a maximum energy of $10^{21}$ eV \cite{Wolfendale}.  
With the Southern
Auger Observatory (3000 km$^2$) a ratio as small as $10\%$ could be
explored at 10$^{19}$ eV with 3 years of data using this new technique.

Our photon bound is also conservative because we have not taken
into account the interactions of the high energy photons in the
magnetic field of the earth \cite{magnetic}. This has the effect
of converting a single energetic photon into a few lower energy
photons.  As the total number of muons in a shower initiated by
a single photon scales with $E^{1.2}$, the number of muons in a
shower initiated by a single photon exceeds the total number of
muons in the multiple photon showers of lower energy.

{\bf Acknowledgements:} This work was partly supported by a joint
grant from the British Council and the Spanish Ministry of Education
(HB1997-0175), by Xunta de Galicia (XUGA-20604A98), by CICYT
(AEN99-0589-C02-02) and by PPARC(GR/L40892). We thank J. Lloyd-Evans
 and F. Halzen for critical readings of the manuscript. We also
thank the authors of the AIRES and CORSIKA codes.


\begin{figure}
\caption{Density maps of two events in the plane perpendicular 
to the shower axis. Recorded muon densities are shown as circles 
with radius proportional to the logarithm of the density. The detector
areas are indicated by shading; the area increases from white to black as
1, 2.3, 9, 13, 34 m$^{2}$. The position of the best-fit core is
indicated by a star. Selected densities are also marked. The y-axis 
is aligned with the component of the magnetic field perpendicular to the
shower axis.}
\label{events.fig}
\end{figure}


\begin{figure}
\caption{Integral number of inclined events as a function of energy 
for the Haverah Park data set compared to the predictions for iron, 
protons and photon primaries. Here the energy is calculated assuming 
a proton primary. The slope of the assumed primary spectrum 
($E^{-1.75}$) is shown to illustrate the increase of trigger 
efficiency with energy.}
\label{rate.fig}
\end{figure}


\begin{table}
\begin{center}
\begin{tabular}{|lcccccccc|} \hline
MR & \multicolumn{2}{c}{ Zenith ($^{\circ}$)} & RA ($^{\circ}$) & Dec. ($^{\circ}$) & \multicolumn{3}{c}{$\log_{10}(E_{p}/\mbox{eV})$} & $\chi^{2}/\nu$ \\ \hline \hline
14050050 & 65 &$\pm$1.2 & 86.7  & 31.7 & 20.09 & -0.15 & +0.26 & 10.3/10  \\
18731630 & 60 &$\pm$2.3 & 318.3 & 3.0  & 20.06 & -0.03 & +0.03 & 45.8/43  \\
14182627 & 70 &$\pm$1.3 & 121.2 & 8.0  & 19.85 & -0.26 & +0.42 &  4.2/10 \\
19167320 & 72 &$\pm$1.3 & 152.5 & 25.9 & 19.82 & -0.06 & +0.04 & 48.4/40 \\
15301069 & 74 &$\pm$1.2 & 50.0  & 49.4 & 19.78 & -0.05 & +0.06 & 26.7/32 \\
12753623 & 74 &$\pm$2.1 & 304.9 & 17.1 & 19.75 & -0.10 & +0.06 & 17.1/11 \\
12519070 & 70 &$\pm$1.3 & 47.7  & 8.8  & 19.62 & -0.08 & +0.06 & 10.2/13 \\ 
\hline
\end{tabular}
\end{center}
\caption{Zenith angle, arrival direction coordinates and shower energy 
(assuming proton primary) of selected showers with energy 
$> 4\times10^{19}$~eV. MR is the event record number. 
The reported $\chi^{2}$ values refer to the energy fits.}
\label{tab}
\end{table}

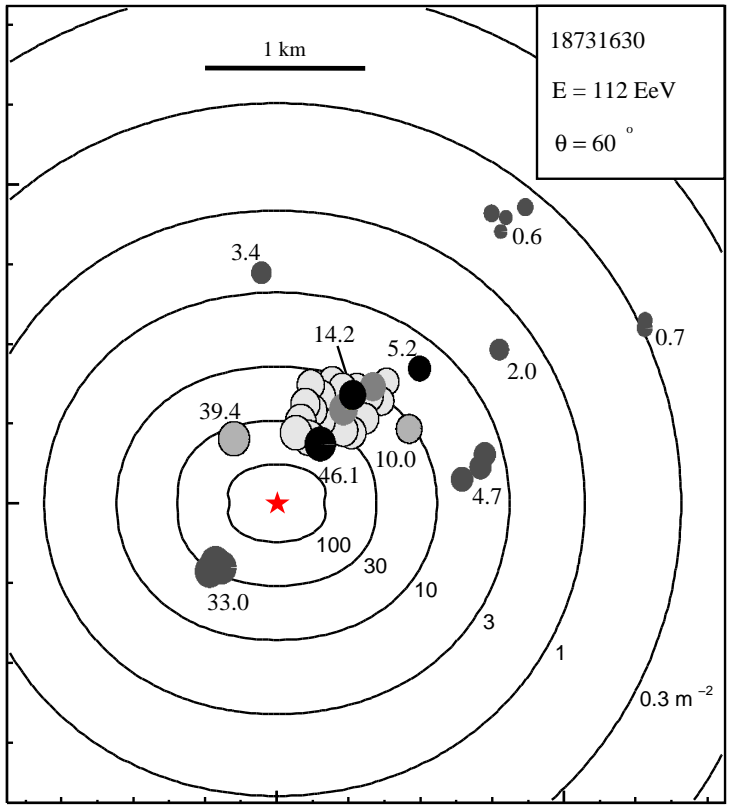
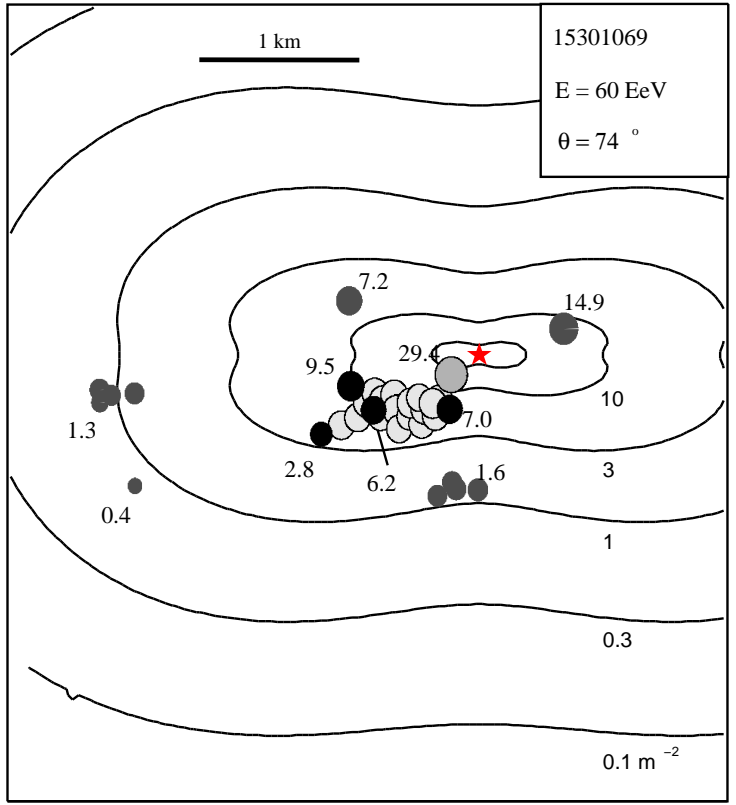
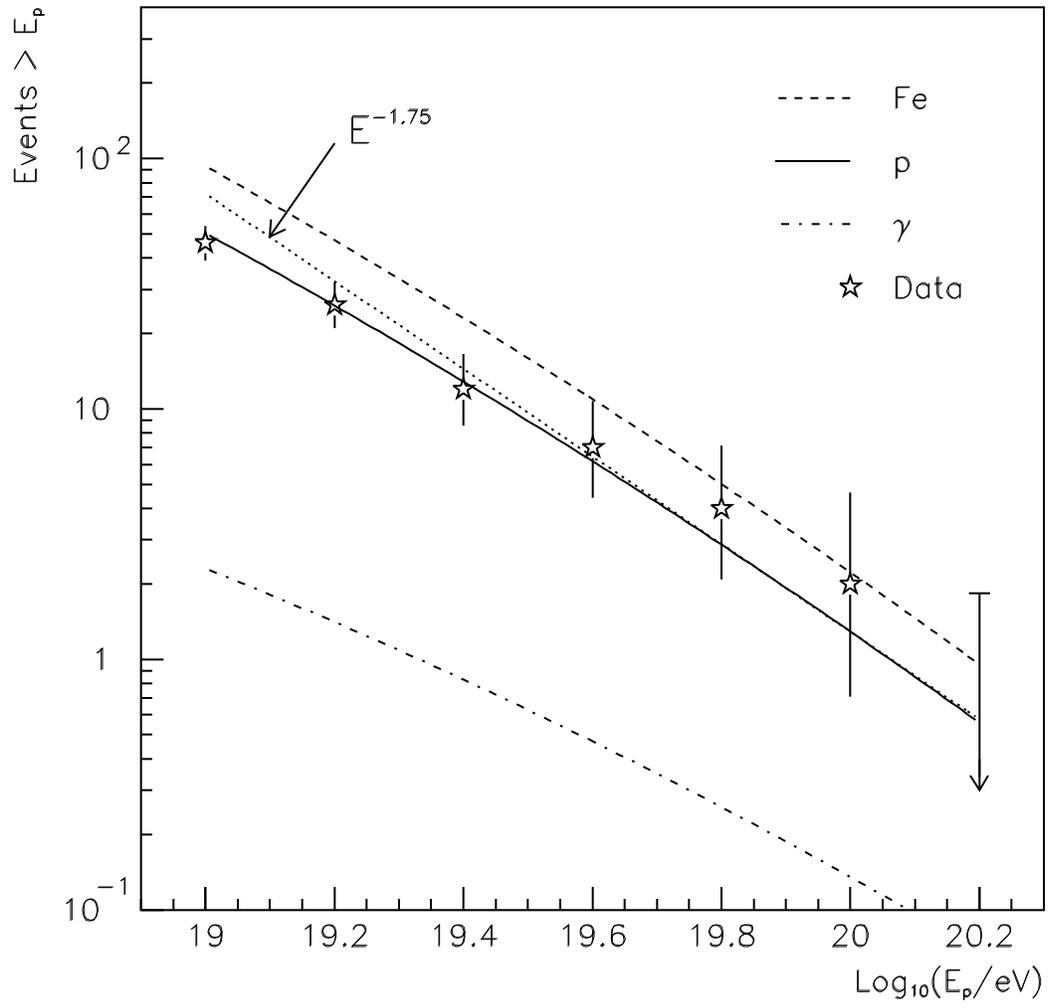
\end{document}